# End-to-End Deep Transfer Learning for Calibration-free Motor Imagery Brain Computer Interfaces

Maryam Alimardani
dept. of Cognitive Science and AI
Tilburg University
Tilburg, The Netherlands
m.alimardani@tilburguniversity.edu

Steven Kocken
dept. of Cognitive Science and AI
Tilburg University
Tilburg, The Netherlands
stevenkocken@hotmail.nl

Nikki Leeuwis
dept. of Cognitive Science and AI
Tilburg University
Tilburg, The Netherlands
n.leeuwis@tilburguniversity.edu

*Abstract*— A major issue in Motor Imagery Brain-Computer Interfaces (MI-BCIs) is their poor classification accuracy and the large amount of data that is required for subject-specific calibration. This makes BCIs less accessible to general users in out-of-the-lab applications. This study employed deep transfer learning for development of calibration-free subject-independent MI-BCI classifiers. Unlike earlier works that applied signal preprocessing and feature engineering steps in transfer learning, this study adopted an end-to-end deep learning approach on raw EEG signals. Three deep learning models (MIN2Net, EEGNet and DeepConvNet) were trained and compared using an openly available dataset. The dataset contained EEG signals from 55 subjects who conducted a left- vs. right-hand motor imagery task. To evaluate the performance of each model, a leave-one-subject-out cross validation was used. The results of the models differed significantly. MIN2Net was not able to differentiate right- vs. left-hand motor imagery of new users, with a median accuracy of 51.7%. The other two models performed better, with median accuracies of 62.5% for EEGNet and 59.2% for DeepConvNet. These accuracies do not reach the required threshold of 70% needed for significant control, however, they are similar to the accuracies of these models when tested on other datasets without transfer learning.

## I. INTRODUCTION

Brain-computer interface (BCI) systems narrow the border between humans and machines by giving the brain the ability to directly control or interact with external devices [1]. Most BCI applications are based on electroencephalography (EEG), which is a non-invasive, easy to use, portable and relatively low cost neuroimaging technique. A commonly used paradigm in BCIs is motor imagery (MI), which relies on the mental process of imagining a movement without actually performing it [1].

However, MI-BCIs are hard to use as they require extensive training until the user is capable of the motor imagery task. Research shows that about 15-30% of users are unable to reach an acceptable level of accuracy (70%) needed for significant control of a MI-BCI application [2-4]. This problem known as BCI inefficiency has limited the widespread adoption of MI-BCIs [5, 6]. To mitigate this problem, past research has either tried to identify the factors that contribute to inter-subject variation in MI-BCI performance [6-8] or focused on development of more powerful AI algorithms that are capable of extracting MI features from EEG signals [9-11].

Recently deep learning (DL) models have gained popularity for classification of EEG signals in brain computer interfaces [12, 13]. Stieger et al. [14] suggested that DL models trained with full scalp EEG could benefit those who struggle with MI-BCI control by extracting useful information from brain areas outside of the motor cortex. A more recent study compared a convolutional neural network (CNN) model that was trained on raw EEG data in an end-to-end approach to a traditional machine learning BCI classifier that relied on MI-specific feature extraction (namely common spatial pattern) [9]. The results indicated an improved classification accuracy in the DL approach for all users although this improvement was significantly larger for the low performers.

Despite their advantages, DL models require large amount of data to be trained and are therefore computationally expensive and time consuming. They can suffer from low classification accuracy when the training dataset is small [15]. This is indeed the case in most MI-BCI experiments, where the number of recorded trials per subject is rather small and there is not sufficient data or time for calibration of the classifier per user. A proposed solution to this problem is subject-to-subject transfer learning (TL) [16]. In BCI applications, TL refers to training a classification algorithm with EEG data from other users in order to transfer the knowledge to a completely new user [16-18]. This method reduces the need for long subject-specific calibration time and hence is promising for future adoption of BCIs outside of the lab.

Previously proposed transfer learning studies have mostly focused on extracting common feature vectors across multiple subjects for the training of a ML classifier [17]. However, deep neural networks can learn and extract features from raw

EEG signals in an end-to-end decoding which removes the need for human-crafted feature engineering steps. This is a rather unexplored terrain in the field of MI-BCIs as only recently DL models gained popularity in EEG classification and more EEG datasets became openly available [9]. Therefore, in this study, we aimed to apply TL in combination with end-to-end DL models to examine whether common features extracted across users can be effectively transferred for subject-independent MI-BCI classification. To achieve this, we chose three state-of-the-art CNN architectures designed for end-to-end EEG decoding (i.e., EEGNet [19], DeepConvNet [20] and MIN2Net [21]) and trained them on a MI-BCI dataset including 55 subjects. The following research questions were formulated:

RQ1: Is there a difference in the performance of EEGNet, DeepConvNet and MIN2Net models, when trained and tested in a subject-to-subject transfer learning approach?

RQ2: Can the combination of TL with end-to-end DL achieve an accuracy above 70% in most participants to surpass the BCI inefficiency threshold in MI classification tasks?

## II. METHODS

### A. Dataset description

The dataset used in this study was gathered by Leeuwis et al. (2021) [22]. The EEG signals were recorded from 55 novice BCI users (19 males, 36 females, $M_{Age}$ = 20.71, $SD_{Age}$ = 3.52). Subjects were all right-handed with (corrected to) normal vision. The EEG signals were recorded from 16 electrodes placed on the sensorimotor area according to the 10-20 system (F3, Fz, F4, FC1, FC5, FC2, FC6, C3, Cz, C4, CP1, CP5, CP2, CP6, T7, T8). During the recording, a 0.5-30 Hz bandpass filter was applied to reduce the noise. The sampling rate was 250 Hz. Each user underwent 4 runs of 40 trials consisting of 20 right- and 20 left-hand motor imagery tasks. The data was stored per run per participant.

### B. Motor Imagery (MI) task

The MI task used during the EEG data collection followed the Graz protocol (Fig.1). In each trial, the participant was first presented with a fixation cross to prepare for the incoming cue (3 seconds). Then they received a cue with a red arrow pointing either to the right or to the left, whereafter they had to imagine moving their corresponding hand/arm (1.25 seconds). The first run was a non-feedback calibration run, where no feedback was given to the subject. In the following three runs, a blue feedback bar was presented, showing the direction and certainty of the classifier's prediction (3.75 seconds). The prediction of the classifier during the feedback stage of the data collection is irrelevant to this study, as here we aim to classify the raw EEG data in an offline manner using deep transfer learning models. In this study, we used the whole duration of the trial for model training and only included the data from feedback runs which equaled 120 trials (3 runs of 40 trials) per subject.

### C. Classification

In this study, three convolutional neural network architectures were compared when trained on the same data.

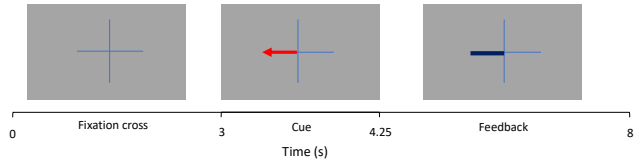

**Figure 2**

*Time course of one trial in the motor imagery task*

Fig. 1. Timeline of one motor imagery trial during EEG data collection [22].

These models were EEGNet, DeepConvNet and MIN2Net, each requiring a specific version of python and tensorflow to run. EEGNet and DeepConvNet can run on python version 3.6 or 3.7 whereas MIN2Net exclusively runs on python version 3.6. All three models need the specific tensorflow version 2.2 and tensorflow-addons version 0.9.1.

*1) EEGNet*: This model is a compact CNN, which was specifically designed for classification of EEG signals [19]. It can be trained with very limited data and can be applied to a variety of different BCI applications. The model consists of 3 convolutional layers; a temporal convolutional layer, designed to learn the frequency filters; a depth-wise convolutional layer to learn frequency-specific spatial filters; and lastly a separable convolutional layer, which combines a depthwise convolution with a pointwise convolution to learn a temporal summary for each individual feature map and optimally mix feature maps together.

*2) DeepConvNet*: This is a model with more generic architecture that can extract a wide range of features to support the idea that CNNs can be used as a general-purpose tool for EEG signal decoding. The model has 4 convolutional layers where the first one is especially designed for EEG signals using spatial filters. The other 3 are standard max-pooling convolutions [20].

*3) MIN2NET*: This model contains three main components that work together. The first component is an Auto Encoder (AE), which uses convolutional layers to create a latent vector (the encoder). Normally autoencoders would then reconstruct the input data from the latent vector (the decoder). However, MIN2Net uses this latent vector and performs deep metric learning on it, which means the model learns to minimize the distance of embedding vectors that have the same label and also maximize the distance for vectors with different labels. At the same time the latent vector is also used as input to a supervised classifier, which uses a SoftMax activation function for classification [21].

### D. Hyperparameter tuning

Hyperparameters can have a big impact on DL models, especially when the wrong values are picked. Table 1 shows a summary of the hyperparameters that were picked for each model. They were carefully selected by observing the effect they had on the model training with the dataset used in this study.

TABLE I. HYPERPARAMETERS FOR EACH DL MODEL

| Model | Epochs | Dropout | LR | min LR | factor | Patience | es_patience |
|---|---|---|---|---|---|---|---|
| EEGNet | 100 | 0.4 | 0.1 | 0.001 | 0.25 | 7 | 18 |
| DeepConvNet | 120 | 0.5 | 0.01 | 0.001 | 0.25 | 10 | 30 |
| MIN2Net | 200 | - | 0.01 | 0.0006 | 0.5 | 12 | 32 |

*Epochs* are difficult to choose, because more epochs never make the model worse, but the number of epochs decide how long the model has to run during training. For EEGNet, the epochs were initially set to 200, but it was quickly found that the model did not improve significantly for epochs between 100 and 200, thus the epochs were set to 100. DeepConvNet improved its loss function significantly less frequent than EEGNet, thus epochs were set to 120, which gave the model more chances to adjust its weights to improve its loss function. MIN2Net improved its loss function even less frequent, therefore the epochs were set to the model's standard of 200 epochs.

The *dropout rate* is a hyperparameter that, during training, randomly selects a given percentage of weights not to include in the training. This is mostly used to prevent overfitting by setting a higher rate for smaller datasets and a lower rate for larger datasets. For EEGNet, the dropout rate was set to 0.4, because it was found that the effect barely changed between 0.3 and 0.5, while values outside this range performed significantly worse. DeepConvNet showed similar findings, also for higher dropout rates of 0.6 and 0.7, thus the dropout rate for this model was set to 0.5. MIN2Net does not have a dropout rate hyperparameter.

The *learning rate (LR)* is the size of the step at which the model adjusts the weights. If the model does not improve its loss function for a certain number of epochs, the learning rate is lowered by multiplying it by the *factor* of the model. By lowering the learning rate, the model can adjust with smaller steps to keep improving more precisely. The number of epochs without improvement of the loss function before the learning rate is lowered is called the *patience* of the model. The *minimum learning rate (min LR)* is the lowest value that the learning rate can be reduced to. For EEGNet, the values of the learning rate, minimum learning rate, factor and patience were set to 0.1, 0.001, 0.25 and 7, respectively. Similar values were set for DeepConvNet, except for patience, which was 10, because DeepConvNet improves its loss function less frequently. For MIN2Net, the values of LR, min LR, factor and patience were set to 0.01, 0.0006, 0.5 and 12, respectively. The factor in this case was set to 0.5, because the patience is high compared to the other models, meaning that the learning rate is adjusted less frequently and hence a larger step is necessary in lowering the learning rate.

*Es_patience* refers to early stopping patience, which is the number of epochs at which the model will stop running if the loss function has not improved. This prevents the model from running too long without improving, making it more efficient. For EEGNet this was not an important hyperparameter, because the model improved relatively often, hence it was set to 18. However, DeepConvNet and MIN2Net improve their loss function less frequently. In this case, a low es_patience means that the models would stop running relatively quickly, yielding a lower classification accuracy for the test subject. This is why the value of es_patience was set higher than EEGNet for both models.

*E. Evaluation*

To evaluate the models using transfer learning, a leave-one-out cross validation approach was taken. This means that in every iteration, the model was trained on the data from 54 (out of 55) subjects and tested on the data from the remaining subject that was not included in the training (data includes 120 MI trial per subject). This process was then repeated for every participant in the dataset. This technique allows to examine whether it is possible to use a pre-trained model to classify MI signals from a new user that was never seen by the classifier before. We report classification accuracies obtained for all subjects as the performance metric of the models. Consequently, we used statistical tests to compare the obtained accuracies across the three models.

### III. RESULTS

Figure 2 demonstrates the distribution of classification accuracies obtained by the selected CNN models for all 55 subjects included in the dataset. EEGNet reached a median test accuracy of 62.5% with the highest accuracy being 97.5%. DeepConvNet produced a median test accuracy of 59.2% with the highest accuracy being 94.2%. MIN2Net reached a median test accuracy of 51.7% with the highest accuracy being 68.3%. As these results suggest, EEGNet performed comparably better than the other two models with MIN2Net showing a median accuracy at the chance level.

To statistically compare the obtained test accuracies across the models, we first evaluated the distribution of values using the Shapiro-Wilk test (EEGNet: $W = 0.93$, $p < 0.001$; DeepConvNet: $W = 0.88$, $p < 0.001$; MIN2Net: $W = 0.94$, $p = 0.008$), which showed that the distribution of accuracies violated the assumption of normality in all models. Therefore, the non-parametric Friedman test was used to compare the three model performances. The test showed that there were significant differences between the three models [$\chi2(3) = 52.77$, $p < 0.001$]. Post-hoc test for pairwise comparison confirmed significant differences between all three models; that is EEGNet could produce subject-independent accuracies that were significantly higher than DeepConvNet ($p = 0.003$) and MIN2Net ($p = 0.001$), and DeepConvNet could significantly outperform MIN2Net ($p = 0.001$) in subject-independent classification.

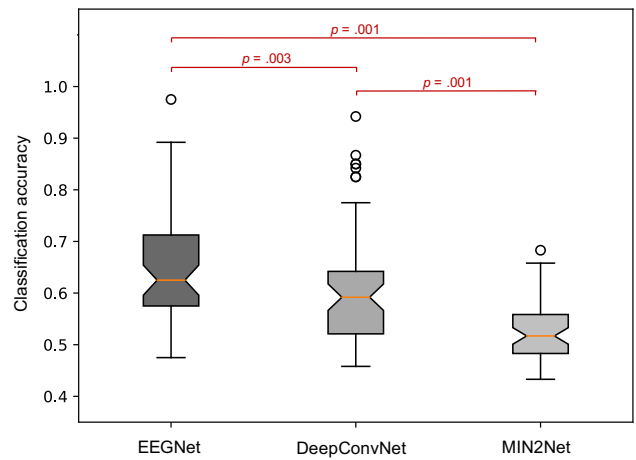

Fig. 2. Comparison of classification accuracies obrained by three DL models using a subject-to-subject transfer learning approach on motor imagery EEG signals from 55 subjects.

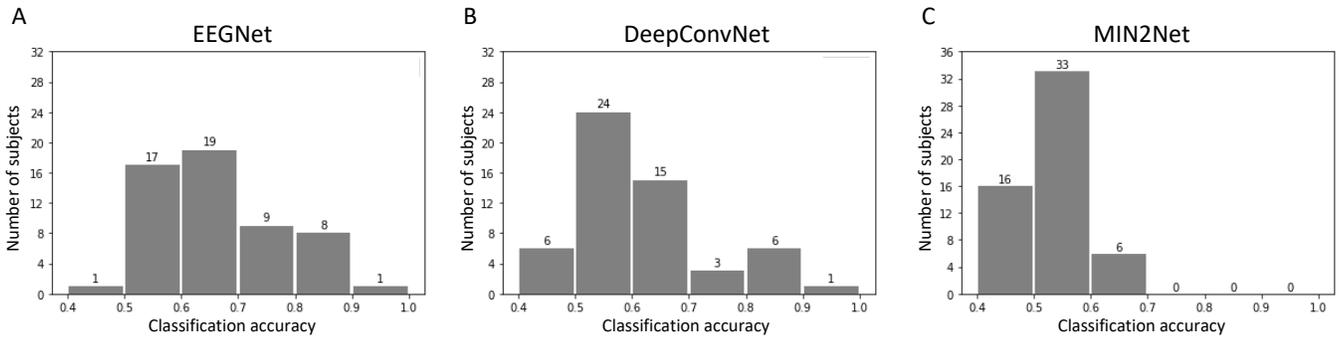

Fig. 3. Histograms indicating distrubution of test accuracies obtained by the three DL models for all 55 subjects.

Figure 3 presents the test accuracy distribution of all 55 subjects for all 3 models. Majority of the subjects classified by EEGNet and DeepConvNet could reach a classification accuracy between 50% and 70% (36 and 39 subjects, respectively), whereas in the case of MIN2Net, almost all subjects had a classification accuracy between 40% and 60%. A BCI user has significant control of a BCI application if an accuracy of 70% or higher is achieved [2, 4, 25]. From Figure 3 it can be inferred that for most subjects, the employed deep transfer learning approach could not yield significant BCI control beyond the threshold. Only 18 subjects predicted by EEGNet and 10 subjects predicted by DeepConvNet achieved accuracies higher than 70%, and none of the subjects classified by MIN2Net passed this threshold.

## IV. DISCUSSION

In order for BCI applications to be more accessible in various domains for general users, they should be easy to use and deliver high and reliable performance. One way to make BCI applications more feasible for usage outside of the lab is to eliminate the calibration time that is needed for the training of the classifier for every new user [17]. To this end, the current study employed a deep transfer learning approach to investigate the efficacy of pre-trained subject-independent DL models to accurately classify MI signals from an unseen user. We trained three CNN architectures using data from 55 subjects in an end-to-end learning pipeline. The results showed that two models, namely EEGNet and DeepConvNet, reached above-chance prediction for the majority of subjects, however, only 18 subjects in EEGNet and 10 subjects in DeepConvNet could yield accuracies beyond the 70% threshold that is required for effective control of MI-BCI applications [2, 24]. Nevertheless, these results are comparable to that of Autthasan et al. [21] who trained the same models employed in this study with three other MI-EEG datasets and obtained similar results.

Among the three CNN models, EEGNet seemed to achieve the highest performance for a subject-independent classification. This can be explained by the design of the model architecture. EEGNet, as the name suggests, is specifically designed to classify EEG data with three convolutional layers, each learning a specific feature from the EEG signals [19]. This is highly specialized as opposed to DeepConvNet [21], which is a more generic architecture that can extract a wide range of features. While DeepConvNet is more versatile for various purposes and datasets, in the case of EEG timeseries, EEGNet proved to perform better.

A possible reason for the low accuracy of MIN2Net is that the designers of MIN2Net aimed to create an efficient model with a low time and computational cost [22]. This was noticeable during the model training; however, it can be argued that such property is incompatible with the advantages of transfer learning. Transfer learning aims to use pre-existing data and train the model to the best of its capabilities, with time mostly not being an issue. The reason for this is that the model only needs to be trained once, whereafter the trained parameters can be transferred over to classify the EEG data of a new user. Trading accuracy for time is therefore unfavorable in this instance.

The novelty of this study mainly lies in the application of end-to-end deep learning models for subject-independent MI-BCI classification. This method leverages information in raw EEG signals that are often lost during signal preprocessing and feature extraction steps in traditional machine learning approaches. Earlier research by Azab et al. [17] and Dehghani et al. [23] showed that it is possible to use subject-to-subject transfer learning to reduce the calibration time, however, as opposed to this study, they employed feature engineering steps that required extensive signal processing. By employing raw EEG signals as input to the CNN models, it is possible extract more discriminative patterns in the data with no human intervention for pre-processing and feature extraction.

While applying deep transfer learning methods in MI-BCI systems can make them more cost-effective and accessible outside of the lab by eliminating the calibration time, they can also negatively affect the system performance for some users and hence worsen the problem of BCI inefficiency. Previous studies showed that DL models can outperform classic machine learning methods, particularly for low performers, when they are trained in a subject-specific manner [9]. However, inter-subject transfer learning faces the challenge of individual variability in neural patterns underlying MI. To mitigate this issue, Zhang et al. [25] proposed finetuning methods that aim to adapt the network parameters for a more robust and generalized recognition of MI patterns.

On the other hand, Pérez-Velasco et al. [26] recently reported a new deep learning architecture (EEGSym), complemented with a data augmentation technique, that could account for inter-subject variability by transferring knowledge across datasets. The results obtained by [26] showed that the model could produce average accuracies above 80% for all datasets and that 95% of the users could reach effective BCI control (>70%). This is contradictory to our results where the model could, at best, yield effective MI-BCI control for only 33% of the users. This difference could be explained based on the TL approach that was used in [26] where data from multiple publicly available datasets was used for the training of the model and instead of holding one subject out, a complete dataset was hold out for testing. This way, it can be argued that the inter-subject variability in one dataset is transferred and accounted for in the other dataset.

Future research could expand this work by employing larger EEG datasets or applying data augmentation methods (as suggested in [26]). Additionally, applying hybrid DL models that combine two different neural networks (e.g., CNN and Long Short-Term Memory (LSTM) [27, 28]) can help gauge individual differences better and improve BCI performance in subject-independent classifications [29]. Finally, an interesting direction would be to investigate the potential of transfer learning in reducing the BCI efficiency problem by transferring knowledge from and across groups of low- and high-performing users (where e.g., performance is determined by conventional ML models during online MI-BCI interaction). Would a low appritude user benefit from knowledge transfer from models trained on high performers' data? Or would it be more beneficial if the models are trained and tested only on low-performing users who might share EEG representations underlying the MI task that are not similar to high performers? [24, 30]. These questions are central to our future research.

V. CONCLUSION

In this study, we proposed an end-to-end deep transfer learning method for subject-independent classification of EEG signals in a left- vs. right-hand motor imagery task. The results indicated that among three state-of-the-art deep learning models, EEGNet could yield the highest accuracy when classifying signals from an unseen subject. Nevertheless, for most subjects, the BCI performance did not reach the required 70% threshold that is needed for significant control. Future research should further expand transfer learning techniques using larger EEG datasets and powerful hybrid models that can sufficiently address inter-subject variability for calibration-free BCI applications.